\documentstyle[preprint,eqsecnum,aps,epsfig,prb]{revtex}

\begin{document}
\draft
\title{High Pressure Study on MgB$_{2}$}
\author{B. Lorenz, R. L. Meng and C. W. Chu$^{1}$}
\address{Department of Physics and Texas Center for \\
Superconductivity, University of Houston,\\
Houston, Texas 77204-5932\\
$^{1}$also at Lawrence Berkeley National Laboratory, 1 Cyclotron Road,\\
Berkeley, California 94720\\
submitted February 14, 2001; revised March 21, 2001}
\date{\today}
\maketitle

\begin{abstract}
The hydrostatic pressure effect on the newly discovered superconductor MgB$%
_{2}$ has been determined. The transition temperature $T_{c}$ was found to
decrease linearly at a large rate of $-1.6$~K/GPa, in good quantitative
agreement with the ensuing calculated value of $-1.4$~K/GPa within the BCS
framework by Loa and Syassen, using the full-potential linearlized augmented
plane-wave method. The relative pressure coefficient, $dlnT_c/dp$, for MgB$_2
$ also falls between the known values for conventional $sp$- and $d$%
-superconductors. The observation, therefore, suggests that electron-phonon
interaction plays a significant role in the superconductivity of the
compound.
\end{abstract}


The recent discovery \cite{akimitsu} of superconductivity in MgB$_{2}$ at
temperatures as high as 40~K has generated great interest. MgB$_{2}$, which
exhibits an AlB$_{2}$ structure with honeycomb layers of boron atoms,
appears to be electrically three-dimensional \cite{kortus} and its grain
boundaries have a far less detrimental effect on superconducting current
transport.\cite{finnemore} The new compound may provide a new way to a
higher superconducting transition temperature $T_{c}$ and an easier avenue
for devices. Two models \cite{kortus,hirsch} were subsequently advanced to
account for the observation. While both have attributed the
superconductivity observed mainly to the conduction bands derived from the
boron sublattice, they propose different mechanisms responsible for the
superconducting pairing. Based on band calculations, Kortus {\it et al.} 
\cite{kortus} suggest that it results from the strong electron-phonon
interaction and the high phonon frequency associated with the light boron
element. A relatively large boron isotope effect on $T_{c}$ has recently
been observed,\cite{budko} consistent with the suggestion. However, Hirsch 
\cite{hirsch} offers an alternate explanation in terms of his ``universal''
mechanism, conjecturing that superconductivity in MgB$_{2}$, similar to that
in cuprate superconductors, is driven by the pairing of the heavily dressed
holes in bands that are almost full to gain enough kinetic energy to
overcome the Coulomb repulsion. A positive pressure effect on $T_{c}$ has
also been predicted by Hirsch if the pressure can reduce the B-B intraplane
distance. We have therefore decided to determine the hydrostatic pressure
effect on $T_{c}$. The $T_{c}$ was found to decrease linearly and reversibly
with pressures at a relatively large rate of $dT_{c}/dP \sim -1.6$~K/GPa up
to 1.84~GPa. The observation is in good quantitative agreement with the
ensuing calculated result of $-1.4$~K/GPa by Loa and Syassen, \cite{loa}
using the full-potential linearized augmented plane-wave method. The
observed value of $dlnT_{c}/dP$ also falls within those of conventional $sp$%
- and $d$-superconductors. The results therefore suggest that
electron-phonon interaction plays a major role in the superconductivity of
this compound.

Polycrystalline MgB$_{2}$ samples examined in the present study were
prepared by the solid-state reaction method.\cite{budko} Small Mg chips
(99.8\% pure) and B powder (99.7\%) with a stoichiometry of Mg:B = 1:2 were
sealed inside a Ta tube in an Ar atmosphere. The sealed Ta ampoule was in
turn enclosed in a quartz tube. The ingredients were heated slowly up to 950 
$^{\circ }$C and kept at this temperature for 2 hours, followed by
furnace-cooling to room temperature. The samples so-prepared were granular
and porous and were used for measurements without further treatment. The
structure was determined by powder X-ray diffraction (XRD), using a Rigaku
DMAX-IIIB diffractometer. The resistivity was determined by the standard
four-lead technique and the $ac$ magnetic susceptibility by an inductance
method with a Linear Research Model LR-700 Bridge. The $dc$ magnetization
was measured using a Quantum Design SQUID magnetometer. The thermoelectric
power was determined employing a home-made apparatus using a sensitive $ac$
measurement technique. The hydrostatic pressure environment was generated
inside a Teflon cell filled with 3M Fluorinert FC 77 acting as the fluid
pressure medium and housed in a Be-Cu high pressure clamp.\cite{chu} The
pressure was estimated using a Pb-manometer situated next to the sample. The
temperature was measured by a chromel-alumel thermocouple located next to
the sample above $\sim 45$~K and by a germanium thermometer housed at the
bottom of the high pressure clamp below $\sim 45$~K.

The powder XRD pattern of the samples displayed the hexagonal MgB$_{2}$
phase but with a very weak trace of MgO. The deduced lattice parameters are $%
=3.084$~\AA\ and $c=3.523$~\AA\, in excellent agreement with the powder
diffraction database.\cite{powder}

The Seebeck coefficient ($S$) of MgB$_{2}$ is positive and relatively small
and decreases with decreasing temperature, as shown in Fig.~\ref{ST},
similar to a metal with effective hole-type carriers. It also exhibits a
rapid drop at 38.9~K and vanishes at 38.1~K, signaling the appearance of a
narrow superconducting transition and consistent with the electrical and
magnetic results to be described below. The temperature dependence of the
resistivity ($\rho $) is shown in the inset to Fig.~\ref{ST}. It decreases
like a metal on cooling, with a resistivity-ratio $\rho $(300~K)/$\rho $%
(40~K) $\sim 3$, much smaller than the $\sim 20$ reported.\cite{finnemore}
We attributed the resistivity-ratio difference to the porosity and the grain
boundary effect of our samples. The $\rho $ starts to drop rapidly at $\sim
39$~K with a rather narrow transition of 0.35~K, defined as the difference
between the temperatures at 10\% and 90\% drops of $\rho $(40 K). Shown in
another inset to the same figure is the $dc$ magnetic susceptibility ($%
\chi_{dc}$) of the sample, measured at 50~Oe as a function of temperature,
in both the zero-field-cooled (ZFC) and the field-cooled (FC) modes. The ZFC-%
$\chi_{dc}$ shows a sharp superconducting transition starting at $\sim 38.5$%
~K with a width of $\sim 1$~K and with more than 100\% superconducting
shielding at 5~K prior to correction of the demagnetization factor. Similar
to ZFC-$\chi _{dc}$, FC-$\chi _{dc}$ demonstrates unambiguously a
diamagnetic shift at $sim 38.5$~K, but the magnitude of the signal is only $%
\sim 1$\% of that for the ZFC-$\chi _{dc}$. This is ascribed to the granular
nature of the sample and the possible strong pinning of the compound.

To determine the pressure effect on $T_{c}$, we chose to measure the $ac$
magnetic susceptibility ($\chi _{ac}$) of the sample in a peak-to-peak field
of $\sim 2$~Oe. At ambient pressure, similar to $\chi _{dc}$, $\chi _{ac}$
undergoes a drastic diamagnetic shift with an onset temperature at $\sim
38.5 $~K, characteristic of a superconducting transition with a mid-point
temperature of $\sim 37.4$~K, as shown in Fig.~\ref{XT}. Under pressure, the
superconducting transition is shifted toward a lower temperature. The
pressure effect on $T_{c}$ is summarized in Fig.~\ref{TP}. It is evident
that $T_{c}$ is suppressed reversibly and linearly at a rate of $%
T_{c}/dP=-1.6$~K/GPa up to 1.84~GPa. The numbers in the figure represent the
sequential order of the experimental runs.

According to the BCS theory, $T_{c}\propto \omega \exp \{-1.02(1+\lambda
)/[\lambda (1-\mu ^{\ast })-\mu ^{\ast }]\}$, where $\omega $ is the
characteristic phonon frequency, $\mu ^{\ast }$ the Coulomb repulsion, and $%
\lambda $ the electron-phonon interaction parameter,\cite{mcmillan} which is
equal to $N(0)<I^{2}{>}/M{<}\omega ^{2}{>}$ with $N(0)$ being the density of
states at the Fermi energy, ${<}I^{2}{>}$ the averaged square of the
electronic matrix element, $M$ the atomic mass, and ${<}\omega ^{2}{>}$ the
averaged square of the phonon frequency. The relative pressure effect on $%
T_{c}$ is $dlnT_{c}/dP=dln\omega /dP+1.02/[\lambda (1-\mu ^{\ast })-\mu
^{\ast }]^{2}(d\lambda /dP)$. Recent band calculations by Kortus {\it et al.}
\cite{kortus} showed that MgB$_{2}$ is electronically isotropic, the $N(0)$
derived mainly from the B atoms near the Fermi surface is large, and the
phonon frequency is high due to the low mass of B, resulting in a large $%
\lambda $. Pressure is expected to increase $\omega $, broaden the density
of states and it may reduce $N(0)$ resulting in a relatively strong decrease
in $T_{c}$. Following the high pressure experiment, Loa and Syassen \cite
{loa} as well as Vogt et al.\cite{Vogt} carried out the full-potential
linearized augmented plane-wave calculation. Loa and Syassen found that MgB$%
_{2}$ is isotropic both electronically and mechanically and found that
pressure suppresses $N(0)$ with $dlnN(0)/dP=-0.31$\%/GPa and enhances $%
\omega $ with $dln\omega /dP=+0.71$\%/GPa. By assuming $\mu ^{\ast }$ and $I$
to be pressure-independent and by adopting the usual numerical values $\mu
^{\ast }=0.1$ and the zero-pressure $\lambda =0.7$, they obtained within the
BCS framework $dlnT_{c}/dP\sim -3.6$\%/GPa or $dT_{c}/dP\sim -1.4$~K/GPa for 
$T_{c}=39$~K. The calculated value may be considered as a crude estimate,
however, it is in good quantitative agreement with our measured $%
dT_{c}/dP=-1.6$~K/GPa. Vogt et al. calculated a similar pressure coefficient 
$dlnN(0)/dP=-0.4$\%/GPa and argued that the pressure effect on $T_{c}$ can
be explained within the BCS- theory and their band structure calculations,
assuming reasonable parameters for $\mu ^{\ast }$ ($0.1$) and $\lambda $ ($%
1.0$). It is interesting that in both calculations the pressure induced
change of $N(0)$ is relatively small compared with the estimated increase of 
$\omega $ indicating that the main source of the decrease of $T_{c}$ with
pressure is its effect on $\omega $.

It has also been demonstrated \cite{levy} that, within the framework of the
BCS theory, the volume effect on $T_{c}$ can be expressed as $%
ln(T_{c}/\omega )/dV\equiv \phi ln(\omega /T_{c})$, where $\omega $ is the
phonon frequency, $V$ the volume, and $\phi $ a material dependent
parameter. For $sp$-superconductors, $\phi \sim 2.5$, while for the $d$%
superconductors, $\phi <2.5$ and can become negative. The lack of knowledge
on $\phi $ and on the compressibility of MgB$_{2}$ prevents us from making a
direct comparison between our observation and the predicted $\phi $.
However, by examining all available data on the relative pressure effect on
the $T_{c}$ of conventional low temperature noncuprate superconductors,\cite
{levy,brandt,smith} we found that, in general, $dlnT_{c}/dP<-8\times 10^{-2}$%
GPa for $sp$-superconductors, but $>-2\times 10^{-2}$/GPa for the $d$%
superconductors, and the value is not sensitive to impurity except for cases
where the Fermi surface topology changes due to applied pressure or impurity
content. For MgB$_{2}$, $dlnT_{c}/dP\sim -4.2\times 10^{-2}$/GPa, which lies
between the values for the two groups of conventional superconductors. It is
interesting to note that $dlnT_{c}/dP\sim -5\times 10^{-1}$/GPa for K$_{3}$C$%
_{60}$,\cite{schirber} in which electron-phonon interaction is considered to
play an important role.

In an alternate approach, regarding the cuprate high temperature
superconductors, Hirsch \cite{hirsch} proposed that MgB$_{2}$ is a
hole-doped superconductor with a conduction band almost completely filled.
The $T_{c}$ varies with carrier concentration non-monotonically and peaks at
an optimal doping level. Pressure is expected to enhance the $T_{c}$
resulting from the reduction of the B-B intraplane distance. Unfortunately,
we found that the $T_{c}$ of MgB$_{2}$ is greatly suppressed by pressure
even though MgB$_{2}$ is mechanically isotropic \cite{loa} and B-B
intraplane distance is expected to decrease under pressures. It should be
noted that a negative pressure coefficient is possible only if pressure can
induce a large change in the carrier concentration and MgB$_{2}$ is
overdoped. The positive $S$ observed by us appears to be consistent with the
hole-doped scenario of MgB${2}$ suggested, although Hall data and the doping
state are still unavailable.

In conclusion, the $T_{c}$ of MgB$_{2}$ has been found to decrease linearly
and reversibly up to $1.84$~GPa at a large rate of $-1.6$~K/GPa, in good
quantitative agreement with the values based on band calculations by Kortus 
{\it et al.} and Loa and Syassen within the BCS framework. The large
relative pressure effect on $T_{c}$ of MgB$_{2}$ also falls within those of
the conventional $sp$- and $d$-superconductors. The observation favors the
proposition that electron-phonon interaction plays a significant role in the
superconductivity in this compound. Unless the pressure can induce a large
hole-transfer in a possibly overdoped MgB$_{2}$ to compensate for the
predicted positive pressure effect on $T_{c}$, the ``universal'' mechanism
cannot account for the observation.

\acknowledgments The authors wish to thank J.~Cmaidalka, J.~Lenzi, and
Y.~S.~Wang for assistance in sample synthesis, thermoelectric power, and
magnetic measurements, respectively. This work was supported in part by NSF
Grant No. DMR-9804325, the T.~L.~L.~Temple Foundation, the John J.~and
Rebecca Moores Endowment and the State of Texas through the Texas Center for
Superconductivity at the University of Houston; and at Lawrence Berkeley
Laboratory by the Director, Office of Energy Research, Office of Basic
Energy Sciences, Division of Material Sciences of the U.~S.~Department of
Energy under Contract No. DE-AC03-76SF00098.

\begin{figure}[tbp]
\caption{$S$ {\it vs.} $T$ of MgB$_{2}$. Upper left inset: $\protect\rho$ 
{\it vs.} $T$. Lower right inset: $\protect\chi_{dc}$ {\it vs.} $T$.}
\label{ST}
\end{figure}

\begin{figure}[tbp]
\caption{$\protect\chi_{ac}$ {\it vs.} $T$ at various pressures.}
\label{XT}
\end{figure}

\begin{figure}[tbp]
\caption{$T_{c}$ {\it vs.} $P$. The numbers represent the sequential order
of the experimental runs.}
\label{TP}
\end{figure}


\begin{references}
\bibitem{akimitsu}  J. Akimitsu, Symposium on Transition Metal Oxides,
Sendai, Japan, January 10, 2001.

\bibitem{kortus}  J. Kortus {\it et al.}, cond-mat/0101446, 30 January 2001.

\bibitem{finnemore}  D. K. Finnemore {\it et al.}, cond-mat/0102114, 6
February 2001.

\bibitem{hirsch}  J. E. Hirsch, cond-mat/0102115, 8 February 2001.

\bibitem{loa}  [Preprint received after initial submission of our 
manuscript] I. Loa and K. Syassen, cond-mat/0102462, 26 February 2001.

\bibitem{Vogt}  [Preprint received after initial submission of our 
manuscript] T. Vogt, G. Schneider, J. A. Hriljac, G. Yang, and J. S.
Abell, cond-mat/0102480, 27 February 2001.

\bibitem{budko}  S. L. Bud'ko {\it et al.}, cond-mat/01014634, 3 February
2001.

\bibitem{chu}  C. W. Chu and L. R. Testardi, Phys. Rev. Lett. {\bf 32}, 766
(1974).

\bibitem{powder}  Powder Diffraction File, Set 38, p. 509 (JCPDS, 1988).

\bibitem{mcmillan}  W. L. McMillan, Phys. Rev. {\bf 167}, 331 (1968).

\bibitem{levy}  M. Levy and J. L. Olsen, Physics of High Pressures and
Condensed Phase, Ch. 13 (Amsterdam: North Holland, 1964).

\bibitem{brandt}  N. E. Brandt and N. I. Ginzburg, Sov. Phys.-Uspekhi {\bf 8}
202 (1964); {\it ibid.} {\bf 12}, 344 (1969).

\bibitem{smith}  T. F. Smith, AIP Conf. Proc. {\bf 4}, 293 (1992); J. Low
Temp. Phys. {\bf 6}, 171 (1972).

\bibitem{schirber}  G. Sparn {\it et al.}, Science {\bf 252}, 1839 (1991).
\end{references}
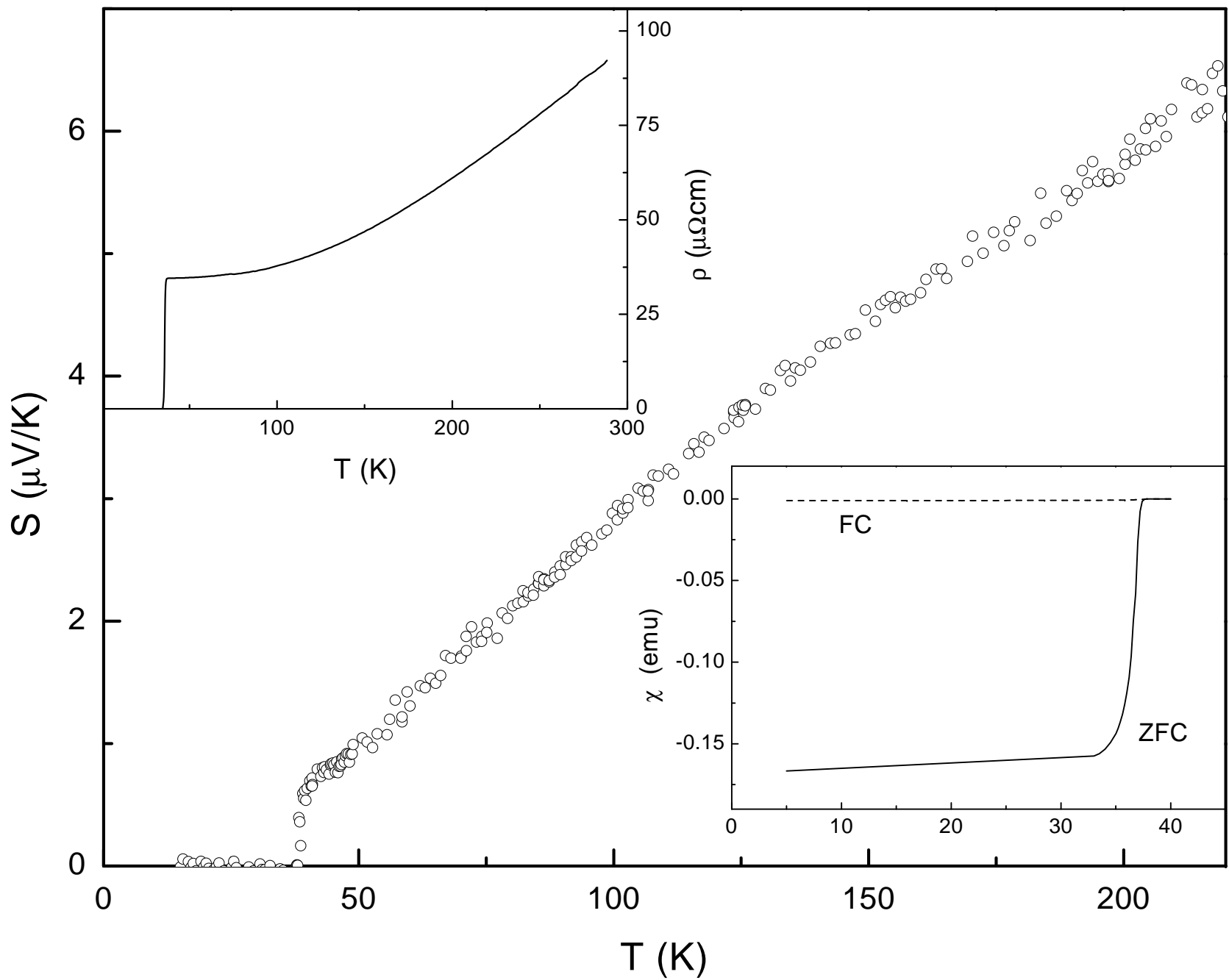
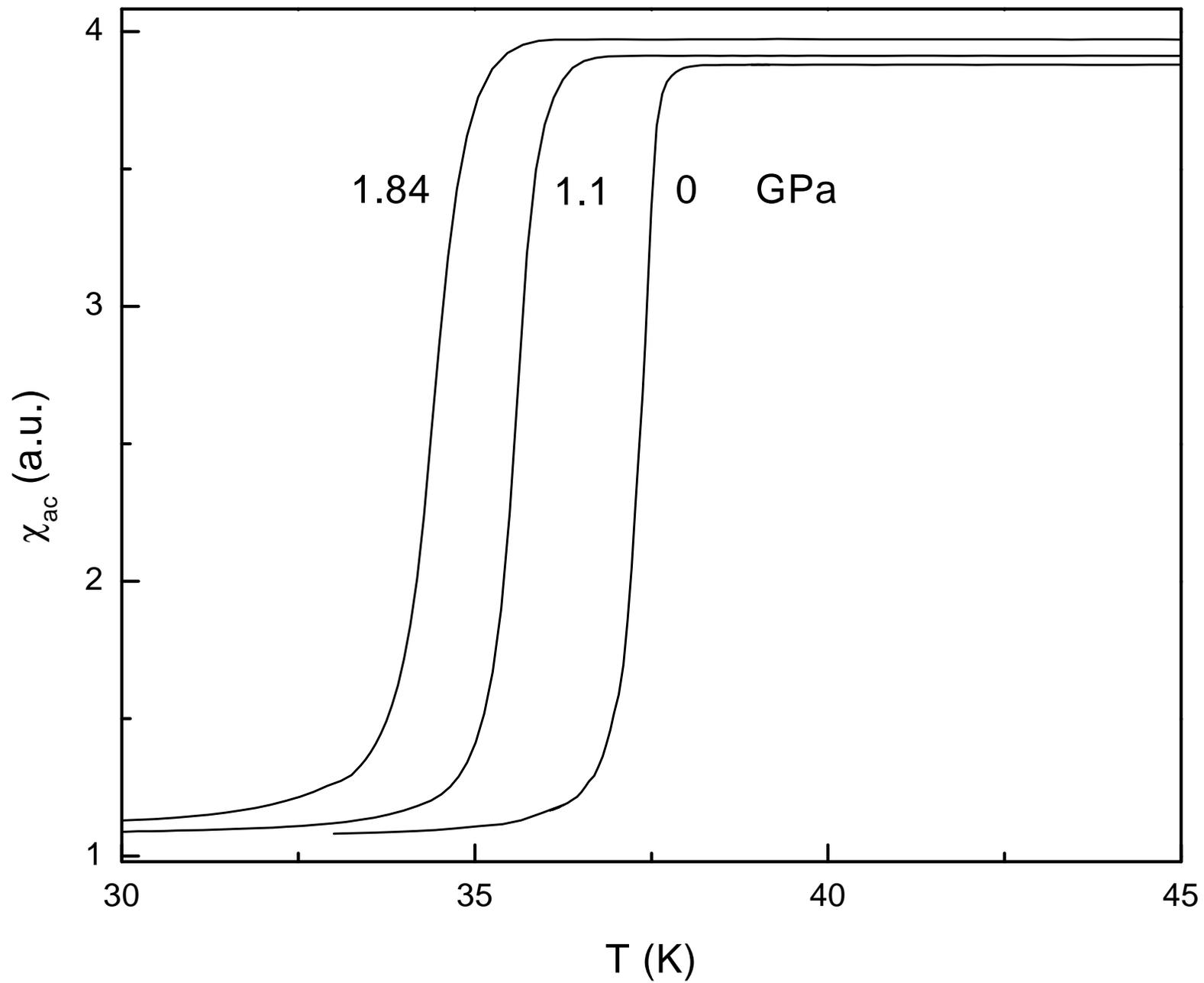
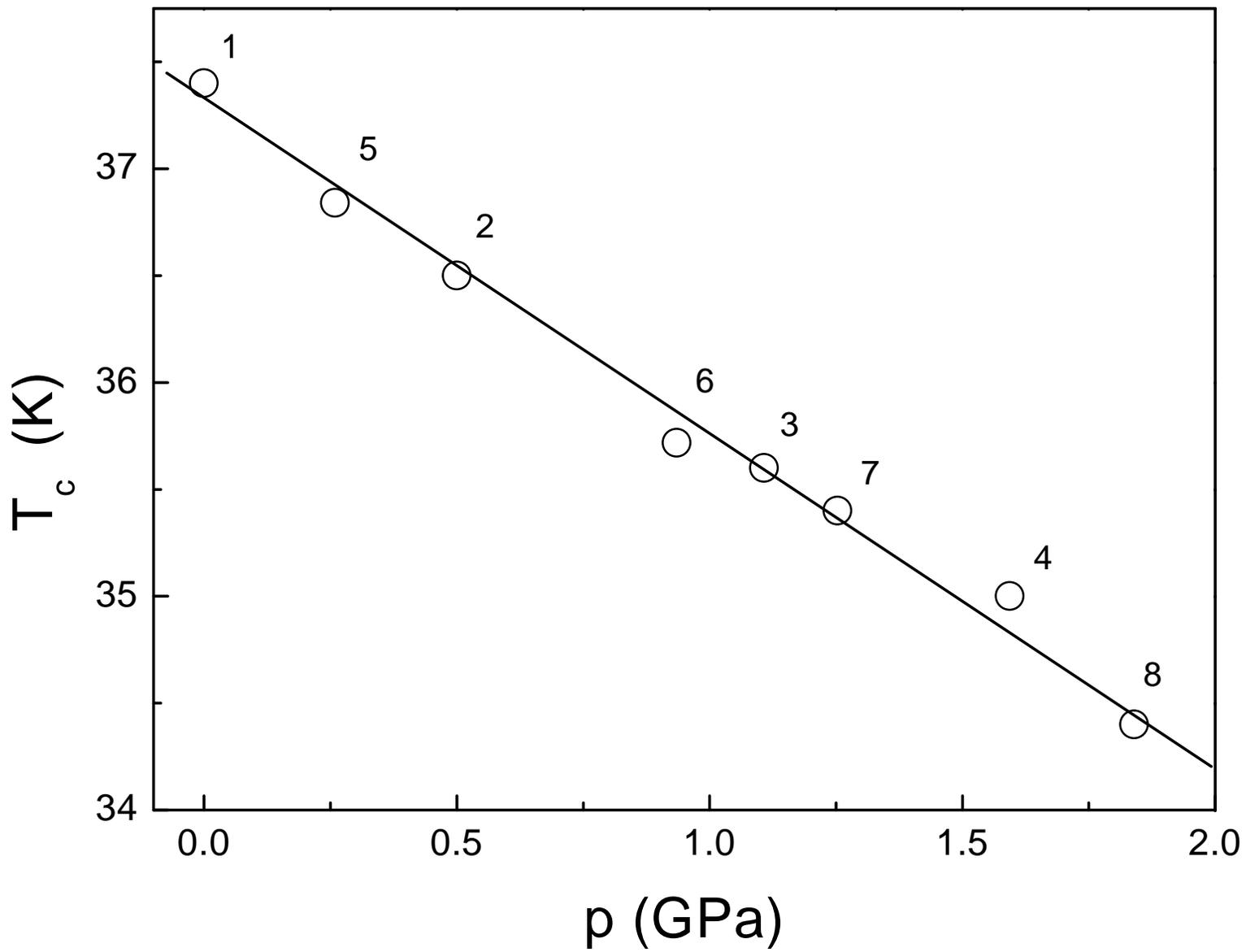
\end{document}